\documentstyle[aps,preprint,epsfig,feynmp]{revtex}

\def\be{\begin{eqnarray}}
\def\ee{\end{eqnarray}}

\newcommand{\bea}{\begin{eqnarray}}
\newcommand{\eea}{\end{eqnarray}}       

\draft 
\begin{document}

\title{Enhanced Minijet Production in $A-A$ Collisions from 
Gluons with Large Transverse Momenta }

\author{H.-J. Pirner and Feng Yuan}
\address{Institut f\"ur Theoretische Physik, Universit\"at Heidelberg, 
Philosophenweg 19, 69120 Heidelberg, Germany}

\date{January 2001}
\maketitle

\begin{abstract}
We find supersaturation for  the intrinsic gluon distribution of nuclei, 
i.e. the low $x$ unintegrated nuclear gluon distribution 
peaks  at intermediate transverse momenta $k_t=Q_s$ and vanishes at zero $k_t$.
Taking into account the
intrinsic transverse momenta of the gluons and the saturation
scale $p_s$ of the produced gluons,
we calculate the minijet cross section arising from gluon gluon scattering
for RHIC energies.  
For central collisions at $\sqrt{s}=200GeV$ the saturation scale 
$p_s \approx 1.4$ GeV is larger due to intrinsic $k_t$ effects   and 
increases with energy.
Our theoretical results on charged particle multiplicity agree very well with 
the recent experimental data from RHIC.
  
\end{abstract}

\section{Introduction}

With the operation of RHIC\cite{phobos,star,phenix},
heavy ion collisions have entered a new era,
where semihard collisions become more important compared to soft
collisions . Therefore a detailed knowledge of 
intrinsic parton distributions is important for a quantitative understanding 
of heavy quark production and/or cross sections at large transverse momenta. 
Due to the increase of the gluon distribution at small $x$,
a sizeable cross section at mid rapidity comes from the region where the
transverse momenta of partons become comparable to their 
longitudinal momenta. Therefore a $k_t$ factorization scheme is preferred, 
where the intrinsic transverse momenta of partons are included
in the calculation of the minijet cross sections. Especially in nuclei,
intrinsic transverse momenta of gluons can become quite large, since at 
small $x$ the gluon clouds of the individual  nucleons overlap and
interact accumulating large transverse momenta.
A careful inversion of the color dipole nuclear cross section allows to
assess this effect. 

	We differentiate between the gluon saturation scale $Q_s$ due to
the interaction of gluons with one another inside the same 
nucleus \emph{before} the collision and the
saturation scale $p_s$ of produced gluons   
\emph{after} the first collision between the nuclei. Both scales arise due to 
different mechanisms,  but sometimes appear to be of the same magnitude.

	Inside the nucleus supersaturation occurs due to the color 
neutrality of the nucleons. One cannot have gluons at small x and small $k_t$,
because  the nucleons from which they originate are color neutral. This finding
is derived from a careful inversion of the multiple scattering of 
a test color dipole on the nucleus. It is confirmed by the evaluation of
the BFKL equation in the nucleus\cite{Braun}. We will give our derivation of
the unintegrated gluon distribution in nuclei in section II.

	In section III we will use this distribution to calculate the mini
jet cross section $g+g \rightarrow g+g$. Due to the intrinsic transverse
momenta of the gluons the LO QCD cross section is enhanced at a given 
cut off scale. Since the intrinsic transverse momenta depend on the
number of gluon clouds overlapping, we find that for more central collisions
the enhancement is larger.

	Section IV is devoted to the analysis of the gluon system which is
produced after the  first collision. These now on shell gluons are numerous at
small nuclear impact parameters and crowd the impact parameter dependent 
area $F(b)$ of intersection between the two nuclei. 
Concentrating on gluons in the central rapidity region one sees that 
their number $\frac {dN}{dy}$  times their individual size $\pi/p_s^2$ 
cannot exceed the area $F(b)$. This equation comes from an  estimate 
of the number of merging gluons $2 \rightarrow 1 $ of produced gluons
compared to the number of splitting gluons $1 \rightarrow 2 $
\cite{MuellerQiu,Blaizot}.
Since the number of
gluons decreases severely as $1/p_s^2$ and 
the saturation criterion $F(b)\times p_s^2$ increases with
$p_s^2$ one obtains a self consistent determination of the saturation
scale of produced partons $p_s$\cite{eskola}. 
Intrinsic transverse momentum will
increase the number of produced gluons at the same cut off.
As a final result, these effects lead  to a 
higher saturation scale $p_s \approx 1.5 $GeV than the scale found
without intrinsic $k_t$.
In the framework of local parton hadron duality the so derived 
gluon distribution can be directly applied to estimate the number of 
produced charged hadrons. From $e^+ e^-$ collisions 
\cite{Dokshitzer} the conversion factor is known and of
order unity.
In this section, we will also present the prediction on charged particle
multiplicity based on our saturation criterion.
As first pointed in \cite{wangp}, the centrality dependence of 
charged particle multiplicity can be used to test different saturation
models.
And very recently, there have been several studies on this quantity by
using different saturation approaches\cite{esk2,kharzeev}.
We will find the saturation criterion we used in the following 
gives reasonable description on centrality dependence of charged particle
reported by PHENIX collaboration\cite{phenix}.
 
	In section V we give a summary of our results and discuss
the possibility to determine the intrinsic saturation scale $Q_s$
by measuring ``monojets''.

\section{Unintegrated gluon distributions in cold nuclei}

We know that the dipole-proton cross 
section depends on the unintegrated gluon distribution function $f_g(x,k_t^2)$.
Assuming a perturbative $2$ gluon exchange picture one has the following 
equation for a color neutral dipole with transverse size $r$, virtuality $Q^2$
and energy $s=(Q^2+m_{\rho}^2)/x$. We assume $\alpha_s$ fixed, \cite{dipole}
\begin{equation}
\label{dp-p}
\hat\sigma_{d-p}(s,r)=\frac{4\pi}{3}\int\frac{d^2k_t}{k_t^2}[1-e^{i\vec{r}
\cdot \vec{k_t}}]\alpha_sf_g(x,k_t^2).
\end{equation}
Here $f_g(x,k_t^2)$ is the unintegrated gluon distribution of the proton,
which is related to the normal integrated gluon density as
\begin{equation}
xG(x,Q^2)=\int^{Q^2}dk_t^2f_g(x,k_t^2).
\end{equation}

The unintegrated gluon distribution function can be obtained from the
color dipole-proton cross section $\hat\sigma_{d-p}$ in Eq.~(\ref{dp-p}) 
by inversion. Since the cross section contains the color dipole factor
$[1-e^{i\vec{r}\cdot \vec{k_t}}]$,
the inversion is more complicated than a Fourier Bessel transform. 
The energy of the dipole cross section under the integrand depends 
on the virtuality $k_t^2$ as $s=(k_t^2+m_{\rho}^2)/x$.
\begin{eqnarray}
\nonumber
\alpha_s f_g(x,k_t^2)&=&\frac{3}{16\pi^3}k_t\frac{d}{dk_t}
k_t\frac{d}{dk_t}\int \frac{d^2 r}{r^2}J_0(k_tr)\hat\sigma_{d-p}(x,r)\\
&=&\frac{3}{16\pi^3}\int \frac{d^2 r}{r^2} [-k_trJ_1(k_tr)-
\frac{k_t^2r^2}{2}(J_0(k_tr)-J_2(k_tr))]\times\hat\sigma_{d-p}(s,r), 
\end{eqnarray}
where $J_0$, $J_1$, and $J_2$ are Bessel functions.

We can derive the unintegrated gluon distribution 
function in nuclei
by using the same method.  
The cross section of a test color
dipole on the nucleus $A$ depends on the nuclear thickness
function $T(b)$. For a hard sphere nucleus we would have 
$T(b)= 2 \rho_0 \sqrt{R_A^2-b^2}$ with $\rho_0=0.16 fm^{-3}$. 
For the following calculations, however, we always use
realistic thickness functions obtained from a nuclear density
distribution  with
finite thickness. This is important for peripheral collisions.
According to Glauber theory the total nuclear cross section 
has the following form:
\begin{eqnarray}
\label{da1}
\hat\sigma_{d-A}(x,r) &=&\int d^2b \hat\sigma_{d-A}(b,x,r),\\
\hat\sigma_{d-A}(b,x,r) &=& 2(1-e^{-\frac{1}{2}T(b)\hat\sigma_{d-p}(s,r)}).
\end{eqnarray}
The differential cross  section $\sigma_{d-A}(b,x,r)$ 
with respect to the nuclear impact
parameter $b$ is related to the unintegrated 
gluon distribution $f_{g/A}(x,k_t,b)$ in the nucleus at impact parameter $b$.
\begin{equation}
\label{da2}
\hat\sigma_{d-A}(b,x,r)=\frac{4\pi}{3}\int{d^2b} \frac{d^2k_t}
{k_t^2}[1-e^{i\vec{r}\cdot \vec{k_t}}]\alpha_sf_{g/A}(x,k_t,b),
\end{equation}
which yields the normal gluon distribution in the 
nucleus after integration.
\begin{equation}
\label{ga}
xG_A(x,Q^2)=\int^{Q^2}dk_t^2d^2bf_{g/A}(x,k_t,b).
\end{equation}
From Eqs.~(\ref{da1}) and (\ref{da2}), we can derive the unintegrated gluon
density of nucleus A as,
\begin{eqnarray}
\alpha_s f_{g/A}(x,k_t,b)
=\frac{3}{16\pi^3}\int 
\frac{d^2r}{r^2} [-k_trJ_1(k_tr)-
\frac{k_t^2r^2}{2}(J_0(k_tr)-J_2(k_tr))]\times 2(1-e^{-\frac{1}{2}
T(b)\hat\sigma_{d-p}(s,r)}).
\end{eqnarray}

One sees that in the limit of small density the integrated density in the
nucleus would be $A$ times as big as in the proton
(shadowing effects will give some modifications). 
To obtain the unintegrated gluon density of nuclei from the above equations,
we must know the parameterization of the dipole cross section
$\hat\sigma_{d-p}(r)$. Up to now, there are various parameterizations for this
function\cite{dosch,wusthoff,kop}. 
As a rough estimate, we use the following simple 
parameterization,
\begin{equation}
\hat\sigma_{d-p}(r)=C(0,s)r^2.
\end{equation}
We take the energy dependent coefficient $C(0,s)$ from the 
parameterization of Ref.\cite{kop},
\begin{equation}
C(0,s)=\sigma_0(s)/R_0^2(s),
\end{equation}
where the parameterizations for $\sigma_0(s)$ and $R_0(s)$ can be found in
\cite{kop}.
To compensate other effects (e.g. shadowing\cite{shad}), 
we normalize our unintegrated gluon density to
the integrated gluon density according to Eq.~(\ref{ga}) for $Q^2\gg Q_s^2$.
Finally, we get the following results for the double differential 
gluon distribution, (cf. Fig.~1)
\begin{equation}
\label{fg}
f_{g/A}(x,k_t,b)=T(b)\times xG_A(x)\frac{k_t^2}{(Q_s^2)^2}e^{-
\frac{k_t^2}{Q_s^2}},
\end{equation}
where
\begin{equation}
Q_s^2=2T(b)\times C(0,s),
\end{equation}
Note that the intrinsic momentum distribution depends on the 
closeness to the center of the nucleus. The smaller $b$ the more gluon clouds
in the nucleus 
overlap producing a larger gluon transverse momentum.
At central $b=0$ the mean transverse momentum in the gluons increases
with nuclear number $A$ as 
$\langle k_t^2\rangle= 2 Q_s^2\propto A^{\frac{1}{3}}$GeV $^2 $. 
The transverse momentum 
distribution peaks at $k_t^2=Q_s^2 $ and then decreases towards small $k_t$,
vanishing at zero transverse momentum.
We call this dynamical phenomena ``supersaturation''.
A similar behavior for the unintegrated gluon distribution
function of nucleus was also derived by the evaluation of
the BFKL equation in the nucleus \cite{Braun}. 
Supersaturation is quite different from saturation 
where growth of the gluon density slows down at small transverse momenta  
cf. \cite {Mueller}.
The dynamical origin of supersaturation at small $k_t$ is the dipole-nucleus
cross section for large dipoles, which is reduced due to multiple scattering.

\section{Produced gluons after one collision in hot interaction area}

Having the unintegrated gluon density of the nuclei, we calculate the
production of gluons and quarks after one collision. To include the intrinsic
$k_t$ effects, we adopt the phenomenological $k_t$-kick model used 
in previous calculations of hard photon and hadron
production\cite{photon,lihn,wang}. 
Following Ref.~\cite{lihn}, we write the
cross section for hard collision as
\begin{equation}
\label{xsa}
\frac{d\sigma(b,p_T)}{dp_T}=\int d^2k_T S(b;k_T)
\frac{d\sigma(p_T')}{dp_T'},
\end{equation}
where $b$ is the impact parameter of the nucleus-nucleus collision.
The differential cross section  
$\frac{d\sigma(p_T')}{dp_T'}$ is identified as the standard 
LO QCD predictions for 
the $2\rightarrow 2$ processes without intrinsic $k_t$ effects,
where the final produced parton would have the  transverse momentum $p_T'$. 
Due to the kick $k_t$ from
the sum of the transverse momenta for the two incident partons,
the real final transverse momentum $\vec p_T={\vec{p}_T}^{~\prime}+\vec{k}_T$.  
The averaging function $S(b;k_T)$ reflects the intrinsic $k_t$ effects
from the incident partons, which depend on the impact parameter 
$b$ of the two nuclei, and can be obtained from the unintegrated gluon
distribution functions calculated in the previous section.
Since the intrinsic gluon distribution functions of the two nuclei depend 
separately on the locations $b_1$ and $b_2$ of the event relative 
to each nuclear center, we first derive the differential 
averaging function depending on $b_1$ and $b_2$,
from Eq.~(\ref{fg}),
\begin{eqnarray}
\label{sb1}
S(b_1,b_2;k_T)&=&\int \frac{d^2q_{1T}}{\pi}\frac{d^2q_{2T}}{\pi}
\frac{q_{1T}^2}{(Q_{s1}^2)^2}e^{-\frac{q_{1T}^2}{Q_{s1}^2}}
\frac{q_{2T}^2}{(Q_{s2}^2)^2}e^{-\frac{q_{2T}^2}{Q_{s2}^2}}
\delta^{(2)}(\vec q_{1T}+\vec q_{2T}-\vec k_T),
\end{eqnarray}
where $b_1,~b_2$, and $b$ satisfy the relation $\vec b_1-\vec b_2=\vec b$,
and $Q_{s1}^2,~Q_{s2}^2$ have the following forms, 
\begin{eqnarray}
\label{qs}
Q_{s1}^2=2T_A(b_1)\times C(0,s),~~~~Q_{s2}^2=2T_B(b_2)\times C(0,s).
\end{eqnarray}
Since the rest of the integrand 
the differential mini jet cross section $\frac{d\sigma(p_T')}{dp_T'}$ 
does not depend on $b_1,~b_2$, we
can integrate Eq.~(\ref{sb1}) over $b_1$ and $b_2$, and obtain a 
normalized averaging function only depending on $b$,
\begin{equation}
S(b;k_T)=\frac{\int d^2b_1d^2b_2T_A(b_1)T_B(b_2)S(b_1,b_2;k_T)\delta^{(2)}
(b_1-b_2-b)}{\int d^2b_1d^2b_2T_A(b_1)T_B(b_2)\delta^{(2)}(b_1-b_2-b)}.
\end{equation}
The above averaging function
$S(b;k_T)$ is normalized to unity after integration over $k_t$.
In Fig.~2, we plot this quantity as a function of $k_t$ for
different impact parameters $b$. This figure shows that for more 
central collisions (smaller $b$), there is a stronger effect from 
intrinsic gluon transverse momenta $k_t$ .

From the above formulas, we derive the number of gluons produced in the 
first collision of nuclei $A+B$ at impact parameter $b$, applying
an infrared cut off $p_0$ on the produced gluons.
\begin{equation}
N_{AB}(b)=T_{AB}(b)\sigma_{hard}(b,p_0),
\end{equation}
where 
\begin{eqnarray}
T_{AB}(\vec b)&=&\int d^2b_1 T_A(\vec b -\vec b_1) T_B(\vec b_1),\\
\sigma_{hard}(b,p_0)&=&\int_{p_T\ge p_0} dp_Td^2k_TS(b;k_T)
\frac{d\sigma(p_T')}{dp_T'}.
\end{eqnarray}
We find that if 
we neglect the intrinsic $k_t$ effects (setting $S=\delta^{(2)}(k_T)$ in 
the above equation), we 
return back to the results in QCD-improved parton model based on
the collinear factorization approach\cite{eskola}.
However, for high energy nuclear collisions, we can not neglect these
intrinsic $k_t$ effects, which  enhance the cross section 
significantly at a given infrared cut off $p_0$.

\section{Saturation of produced gluons and inclusive cross section 
}

For central collision ($b=0$) of identical nuclei A$+$A with radius $R_A$, 
the interaction area of the two nuclei is
identical to their transverse size $\pi R_A^2$.
Assuming the radius of each produced gluon with transverse momentum $p_0$ 
to be $1/p_0$, we have saturation for an infrared cut off $p_0=p_s$ 
when the produced quanta 
fill the transverse interaction area $\pi R_A^2$.
All gluons with momenta smaller than the saturation momenta fuse into gluons
with a total momentum larger than the saturation momentum. They are not 
realistic endproducts for the application of local parton hadron duality
to the minijet cross section.
So, for central collision, we have the following saturation condition, 
\begin{equation}
N_{AA}(p_0=p_s,b=0)\frac{\pi}{p_s^2}=T_{AA}(0)\sigma_{hard}(p_s) 
\frac{\pi}{p_s^2}=\frac{\pi R_A^2}{\beta}.
\end{equation}
The parameter $\beta$ takes care of some modification of this simple 
geometrical picture which 
can be estimated from the evolution equation
including gluon gluon fusion \cite {MuellerQiu} or from
\cite{Blaizot} and is expected to be around $2-3$.
More realistic may be a simulation of classical Yang-Mills theory
which gives $\beta=1.4-2.0$ \cite{McLerran,Venugopalan}.

We extend the above saturation equation to any arbitrary impact 
parameter $b$ of the two nuclei,
\begin{equation}
\label{sat}
N_{AA}(p_0=p_s,b) \frac{\pi}{p_s^2}=T_{AA}(b)\sigma_{hard}(p_s(b);b) 
\frac{\pi}{p_s(b)^2}=\frac{F(b)}{\beta},
\end{equation}
where $F(b)$ is the interaction area for $b>0$.
From the geometry of the collision at arbitrary impact parameter
$b$ and taking into account the realistic nuclear density profiles of 
two nuclei, we have the following form for the function $F(b)$: 
\begin{equation}
F(b)=\int d^2b_1d^2b_2 (1-e^{-T_A(b_1)\sigma_{in}})
(1-e^{-T_A(b_1)\sigma_{in}})\delta^{(2)}(b_1-b_2-b),
\end{equation}
where $\sigma_{in}$ is the inelastic cross section for proton-proton 
scattering. In our calculations, we use $\sigma_{in}=35,~39,~45$mb 
for $\sqrt{s}=56,~130,~200GeV$, respectively.
In the limit of sharp-edged nuclear density, this area function 
will lead to the normal area function of two incident nuclei 
at impact parameter $b$ 
$$
F(b)\rightarrow 
2\left [ R_A^2 arccos(\frac{b}{2R_A})-\frac{b}{2}\sqrt{R_A^2-b^2/4}
\right ].
$$
Naturally, the saturation momentum becomes a function of the 
impact parameter $p_s=p_s(b)$. 
Having determined the saturation scale $p_s(b)$ from Eq.~(\ref{sat}),
 we plot in Fig.~3 the numerical results for $p_s$ as a 
function of $b$ for three typical energies at BNL RHIC, $\sqrt{s}=56$,
$130$, and $200 GeV$, where we set $\beta=1.8$.
From this figure, we can see that the saturation scale $p_s$ increases with 
collision energy. For central collision ($b=0$), $p_s=1.2GeV$ at 
$\sqrt{s}=56GeV$, increases to $p_s=1.3GeV$ at $\sqrt{s}=130GeV$,
and to $p_s=1.35GeV$ at $\sqrt{s}=200GeV$.
 
It is important to differentiate between the saturation 
scales $Q_s(b_1),Q_s(b_2)$ of intrinsic virtual gluons 
and the saturation scale $p_s(b)$ of produced gluons.
The intrinsic saturation scale is a property of each individual 
cold nucleus and is defined at the local impact 
parameter in each nucleus. It parameterizes the intrinsic 
gluon distribution functions in both nuclei from which 
the averaging function $S(b,k_T)$ is derived. The saturation scale for the
produced gluons $p_s$ is a property of the interacting system 
of the two nuclei and depends on the nucleus nucleus impact parameter $b$.     

Our results show only a weak dependence of $p_s$ on $b$ which
is  not so dramatic as found in ref. \cite{esk2}.
Furthermore, the saturation scale $p_s$ in our model is larger than $1GeV$
for a wide range of $b$.
We note that the intrinsic saturation scale $Q_s$ for the gluon 
distribution in each individual nucleus (\ref{qs}), however, is sensitive 
to the cylindrical distance $b'$ of the center of the nucleus.
In Fig.~3 we show both saturation scales for
comparison.

With the saturation scale $p_s(b)$ determined from Eq.~(\ref{sat})
we can calculate the number of partons produced after the first collision
of $A+B$, from which we can further estimate the charged particle
multiplicity \cite{Dokshitzer,eskola,esk2}. 
In Fig.~4 we plot 
the quantity $\frac{dN_{ch}}{d\eta}/(0.5N_{part.})$ 
as a function of $N_{part.}$. The number of participants $N_{part.}(b)$ 
in an A$+$A collision can be calculated using
\begin{equation}
N_{part.}=2\int d^2b_1T_A(b_1)(1-e^{-T_A(b_2)\sigma_{in}}).
\end{equation}
The charged particle multiplicity $\frac{dN_{ch}}{d\eta}$ is related to the
parton number produced by the first collision $\frac{dN_{ch}}{d\eta}=
\frac{2}{3}\frac{dN}{d\eta}$.
To take into account the difference of rapidity and the pseudo-rapidity, we 
include another factor $0.9$ for 
$\frac{dN}{d\eta}=0.9\frac{dN}{dy}$\cite{eskola,esk2}, 
which can be calculated from above
formulas.
From Fig.~4, we see that our predictions on centrality dependence of
charged particle multiplicity agree well with the PHENIX data 
as well as the PHOBOS data except few data points associated with very
peripheral collisions, for which the saturation mechanism may not
be valid any more.

As a final remark, we note that in \cite{esk2}, quite different from 
our global saturation approach discussed above,
a local saturation condition was introduced 
for every $b_1$ and $b_2$ in each nucleus. If the so determined local 
$p_s(b_1,b_2)$ is getting too small, an additional cut off is implemented
in ref. \cite{esk2} 
such that $p_s(b_1,b_2)\geq 0.5$ GeV. 
This local saturation approach leads to a stronger discrepancy for peripheral
collisions than our model cf. Fig~4 and Fig.~4 of Ref.\cite{phenix}.
However, to finally settle this issue, we need more
experimental observables and data as suggested in\cite{wang3}.

\section{Conclusions}    

In this paper, we have calculated the minijet production in $A-A$ collisions
at RHIC, including the enhancement from intrinsic transverse momenta of 
the incident gluons. 
We derived the unintegrated gluon distributions of cold 
nuclei and found ``super saturation''; the number density of
gluons vanishes at zero transverse momentum and has a maximum at 
the intrinsic saturation scale $Q_s$.
We differentiate between this gluon saturation scale $Q_s$ 
inside both nuclei \emph{before} the collision 
and the saturation scale $p_s$ of produced gluons   
\emph{after} the first collision between the nuclei. 
To obtain the saturation scale $p_s$, we apply the saturation 
criterion in the central rapidity 
region, where 
the produced parton number $\frac {dN}{dy}$ times their individual
size $\pi/p_s^2$ must not exceed the area $F(b)$ of the two interacting
nucleus. 
Intrinsic transverse momentum of the incident partons 
increases the number of produced gluons (i.e., the production 
cross section) at the same cut off, and then
leads  to a higher saturation scale $p_s \approx 1.5 $GeV than the scale found
without intrinsic $k_t$.
Finally, we find that our numerical results on the centrality dependence of the
charged particle multiplicity agree well with the experimental 
PHENIX and PHOBOS
data, which support the saturation scheme we used in this paper.

We note as a final remark, that it is also very important
to directly determine the intrinsic transverse momentum scale 
$Q_s$ for large nuclei.
For this  purpose we propose to measure ``monojet'' production at RHIC , 
in which the final state only contains one large $p_T$ jet
without any other large $p_T$ balancing jets.
By triggering on produced particles in the projectile and target 
fragmentation regions with a summed high transverse
momentum in one azimuthal hemisphere, 
one may select minijets with transverse momentum in the opposite azimuthal 
hemisphere which  come from 
$2 \rightarrow 1 $ processes in the central
region. In principle one can thereby
test the intrinsic saturation scale $Q_s$ which is characteristic for 
isolated nuclei and also accessible in electron nucleus scattering.
The same configuration of monojet production was also suggested 
in\cite{monojet} to search the jet quenching effects in high energy parton 
production.
The difference between these two approaches and the experimental accessible
will be discussed in the future studies.

\acknowledgments
We are grateful for the correspondence with A. Accardi, M. Braun, 
D. Kharzeev and X.N. Wang.
We thank J.~H\"ufner, B. Kopeliovich, C.~Ewerz, and A.~Shoshi 
for discussions.

%
%


\newpage
\vskip 10mm
\centerline{\bf \large Figure Captions}
\vskip 1cm

\noindent
FIG.1. The double differential gluon distribution $f_{g/A}(x,k_t,b)$ for
Au$(A=197,~R_A=6.37~fm)$ at 
$x=10^{-2}$ as a function of $k_t~[$GeV$]$ at different
cylindrical distances ($b=0,~1/2R_A,~3/4R_A,~R_A$) from the central
axis of the nucleus.
The maxima of these distributions are at $k_t=Q_s\approx 0.96GeV,~
0.90GeV,~0.76GeV,~0.46GeV$ for these four cases
respectively.

\noindent
FIG. 2. The averaging function $S(b;k_T)$ vs $k_T$ for different 
nucleus-nucleus impact parameters $b$ shows that intrinsic transverse 
momentum kicks $k_t$ with 
larger $k_T$ are more numerous for central collisions at $b=0$ 
than for peripheral collisions with large $b$.

\noindent
FIG. 3. The saturation scale $p_s$ in $GeV$ is shown as a function 
of $b$ for three
different energies $\sqrt{s}=56GeV,~130GeV,~200GeV$.
We also plot the intrinsic saturation scale $Q_s$ ($x=0.01$) 
as a function
of $b'$ for the individual nuclei before collision, 
where $b'$ is the cylindrical distance from the central
axis of the nucleus. In contrast to $p_s$, $Q_s$  is a more 
strongly varying function of $b'$.

\noindent
FIG. 4. The theoretical normalized charged particle multiplicity  
$[dN_{ch}/d\eta]/(0.5N_{part.})$ as a function
of the number of participants $N_{part.}$ is shown for the theory
with intrinsic gluon transverse momentum $k_t$ (full line) and 
without $k_t$ averaging effects (dashed line).
The crosses mark the experimental PHENIX ($+$) and PHOBOS ($\times$) data.

\begin{figure}[thb]
\begin{center}
\epsfig{file=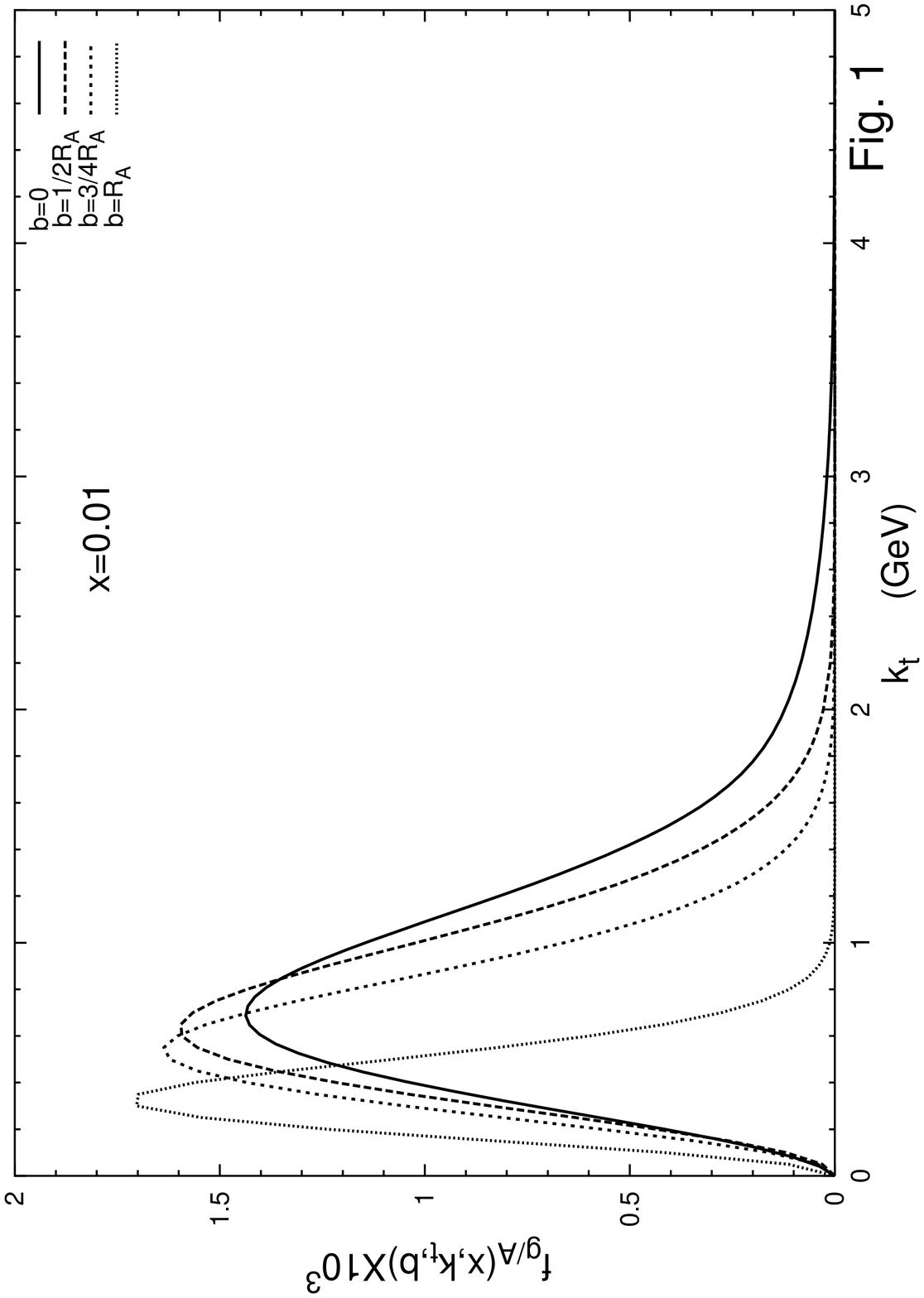,angle=0,width=16cm,height=20cm}
\end{center}
\end{figure}

\begin{figure}[thb]
\begin{center}
\epsfig{file=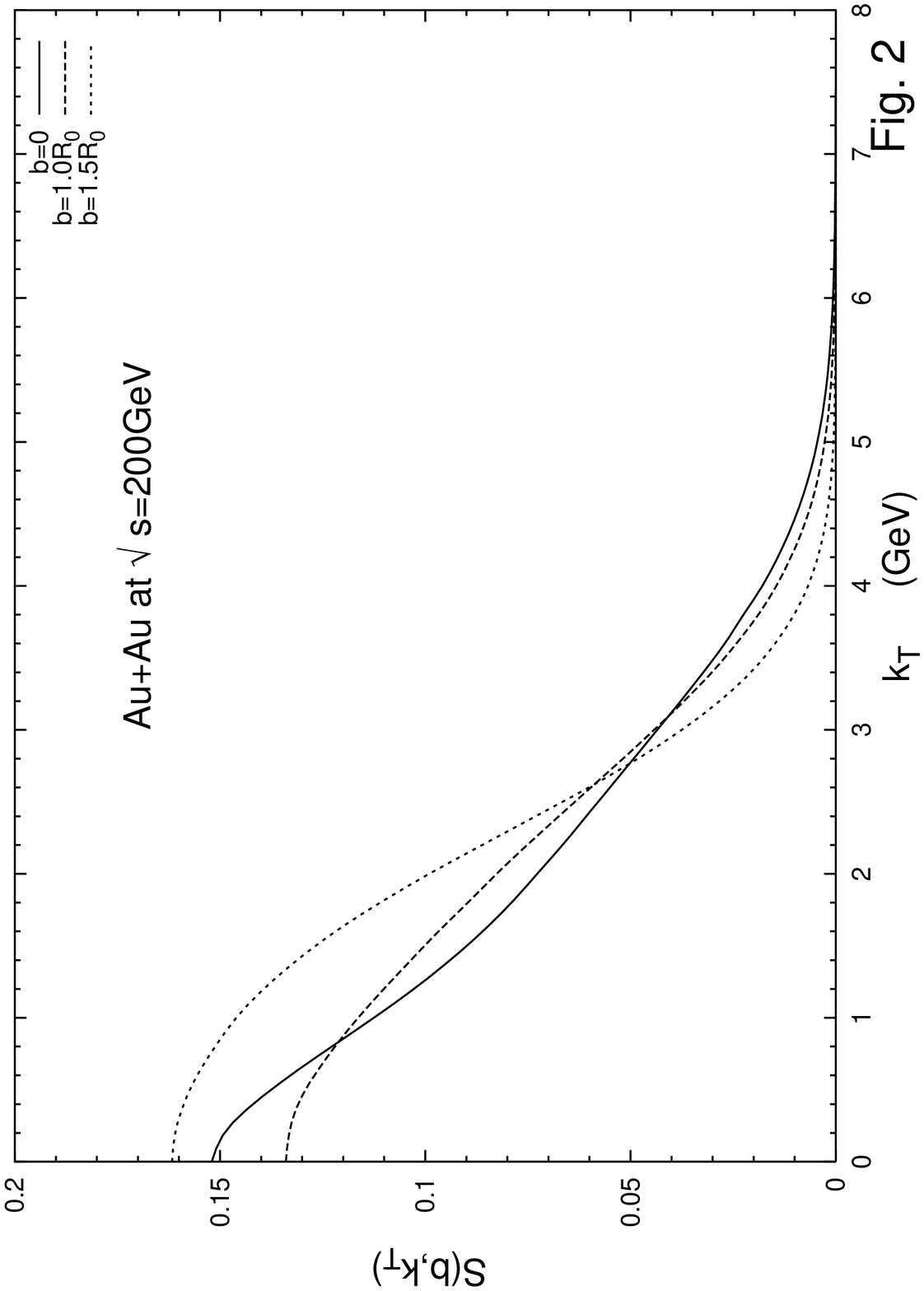,angle=0,width=16cm}
\end{center}
\end{figure}

\begin{figure}[thb]
\begin{center}
\epsfig{file=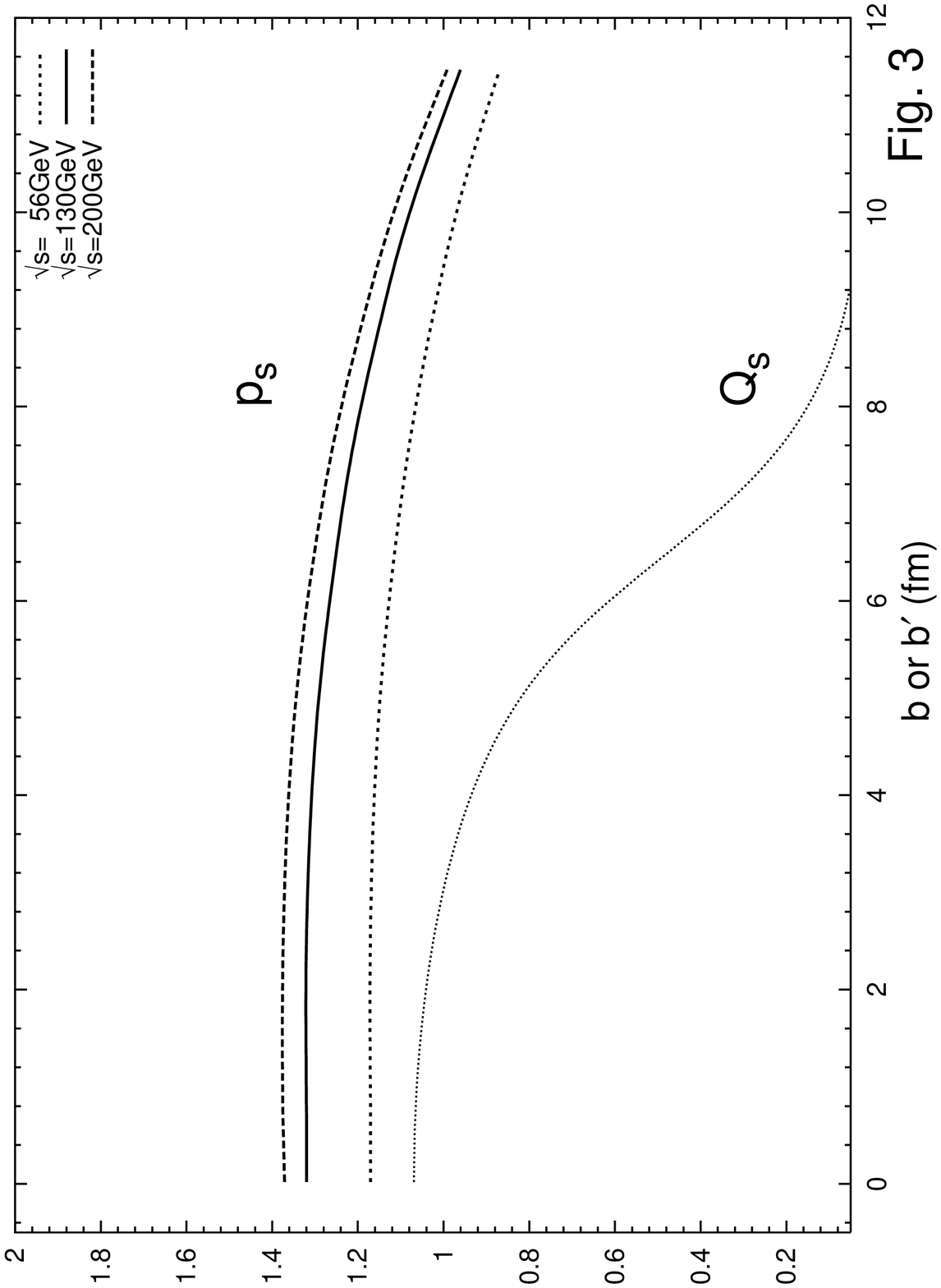,angle=0,width=16cm}
\end{center}
\end{figure}

\begin{figure}[thb]
\begin{center}
\epsfig{file=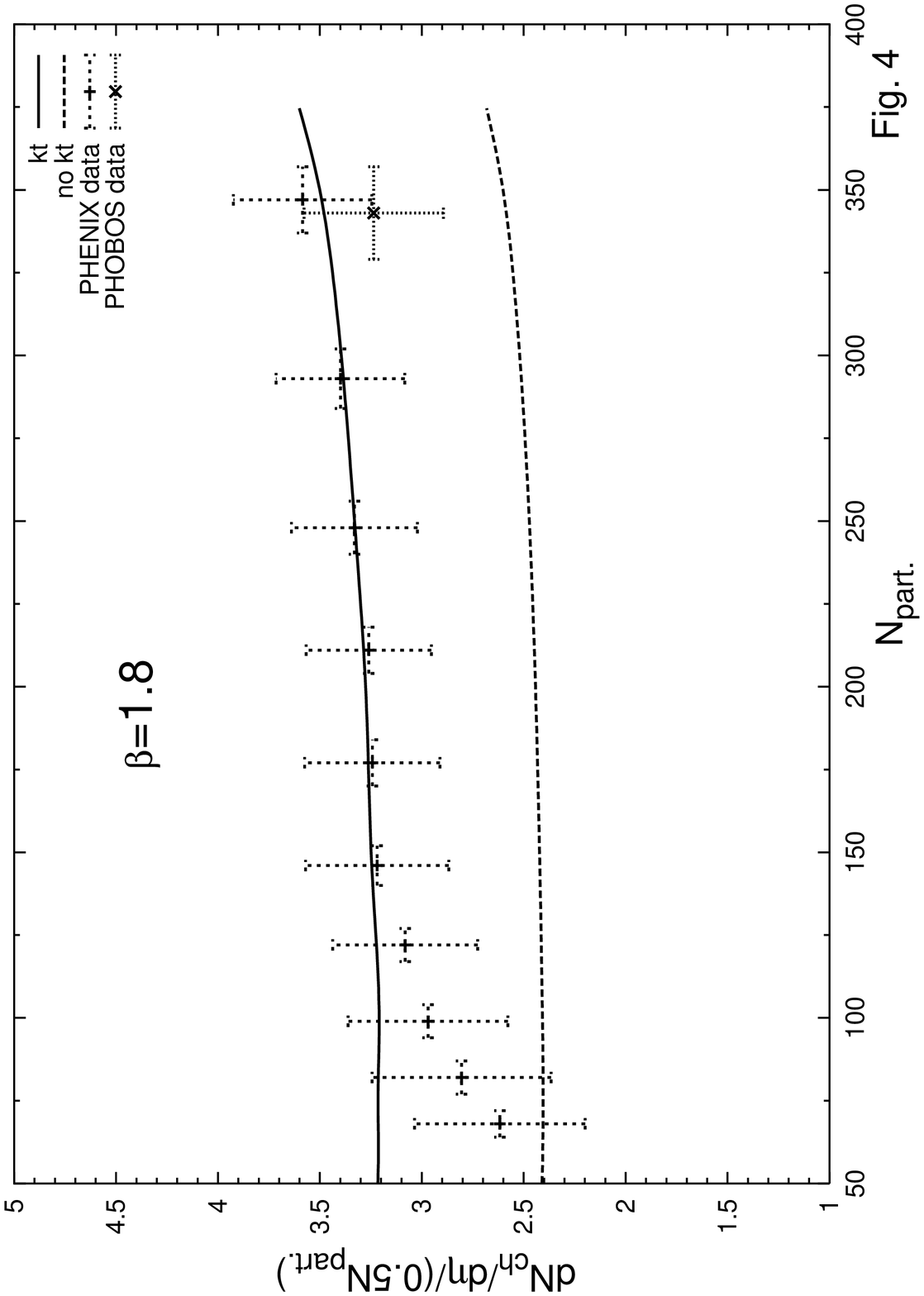,angle=0,width=16cm}
\end{center}
\end{figure}

\end {document}